\documentclass[aps,eqsecnum,showpacs,twocolumn,superscriptaddress]{revtex4}

\usepackage{graphicx}

\usepackage{amsmath}
\usepackage{amsfonts}
\usepackage{amssymb}
 
\usepackage{bm}

\begin{document}

\title{ Theory of 2$\delta$-kicked Quantum Rotors}

\author{C.E.~Creffield}
\affiliation{Department of Physics and Astronomy, University College London,
Gower Street, London WC1E 6BT, UK }

\author{S.~Fishman}
\affiliation{Physics Department, Technion, Haifa, IL-32000, Israel}

\author{T.S.~Monteiro}
\affiliation{Department of Physics and Astronomy, University College London,
Gower Street, London WC1E 6BT, UK }

\date{\today}

\begin{abstract}
We examine the quantum dynamics of cold atoms subjected to 
{\em pairs} of closely spaced $\delta$-kicks from standing waves of light,
and find behaviour quite unlike the well-studied quantum kicked rotor (QKR). 
Recent experiments 
[Jones et al, {\em Phys. Rev. Lett. {\bf 93}, 223002 (2004)}]
identified a regime of chaotic, anomalous classical diffusion. We show that
the corresponding quantum phase-space has a cellular structure, arising from
a unitary matrix with oscillating band-width.
The corresponding eigenstates are exponentially localized, but scale 
with a fractional power, $L \sim \hbar^{-0.75}$, 
in contrast to the QKR for which $L \sim \hbar^{-1}$. 
The effect of inter-cell (and intra-cell) transport is investigated
by studying the spectral fluctuations with both periodic
as well as `open' boundary conditions. 
\end{abstract}


\maketitle

\section{Introduction}
A recent experimental study \cite{Jones} of cesium atoms subjected
to pairs of near $\delta$-kicks using pulsed optical potentials
showed behaviour surprisingly different from the well-studied single-kick
system, the quantum kicked rotor (QKR or $\delta$-KR). The QKR is possibly
the most studied experimental and theoretical
paradigm of classical Hamiltonian chaos. The classical dynamics
of the system make a gradual transition to chaotic dynamics as a
function of an effective kick strength $K$ (related to the intensity of
the optical potential). It has also been well-investigated
experimentally, using mainly cesium atoms  \cite{Raizen}.
In the large $K$ (chaotic) regime, typical 
{\em classical} trajectories are diffusive; 
to lowest order, the diffusion is a random
walk in momentum, with diffusion constant
$D \approx \frac{K^2}{2}$. Then, for a
corresponding  ensemble of atoms, say, the average kinetic energy 
increases linearly with time i.e. $\langle p^2 \rangle/2 \sim Dt$. 
Short-ranged correlations to the classical diffusion do exist, however, 
and their effects have also been observed experimentally \cite{Raizen2}.

In the corresponding quantum system, the quantum kicked rotor (QKR),
the diffusion is arrested on a timescale $t^* \sim D/\hbar^2$.
In experiments and calculations,
an initially gaussian momentum distribution $N(p,t=0)$,
is seen to evolve into
an exponentially localized distribution $N(p, t > t^*) \sim \exp {-|p|/L}$,
where $L \sim \frac{D}{\hbar}$. This quantum suppression of chaotic
diffusion is an important
quantum chaos phenomenon termed Dynamical Localization \cite{Casati}. A formal
connection between Dynamical Localization and Anderson localization,
(the exponential localization of electronic wavefunctions in 
disordered metals) has been made \cite{Fish}.

Here we consider instead
the case where the particles are subjected instead to {\em pairs} 
of closely spaced kicks: the $2\delta$-kicked rotor ($2\delta$-KR).
The recent experimental study \cite{Jones}
showed that the corresponding classical diffusion is anomalous, 
and involves many additional corrections from weak, but long-ranged (in time) 
correlations. The experiment identified periodically spaced
momentum-trapping regions
where the atoms absorb little energy, interspersed by regions where they
absorb energy even more rapidly than the standard $\delta$-KR system.

\begin{figure}[htb]
\includegraphics[width=3in]{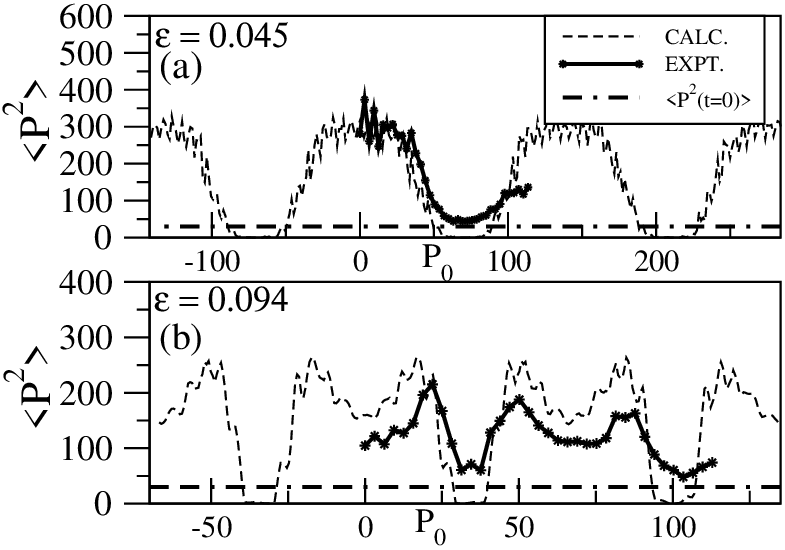}
\caption{Experimental results from \cite{Jones} 
showing effects of momentum trapping.
 Each point indicates the 
energy absorbed by a cloud of cesium atoms after
dynamical localization.
The horizontal-axis gives the initial average
momenta $\langle p \rangle=P_0$ of each cloud of cesium atoms: 
i.e. each atomic cloud initially
has a gaussian momentum distribution well-centred on $P_0$.
The trapping regions correspond to $\langle p^2 \rangle \simeq 0$.
 All the data corresponds to effective values of 
$K=3.2$ and $\hbar=1$. }
\label{Fig1}
\end{figure}

In this work we extend and develop a preliminary study of this system 
\cite{preprint}, and reveal unusual localization properties
and spectral fluctuations not seen in the standard QKR.
We find a novel dynamical localization regime, with generic localization
lengths $L \propto \hbar^{-0.75}$ 
determined by a {\em fractional} exponent ($\simeq 0.75$). 
A similar exponent was found in  
$P(t)=|\langle \psi(t=0) | \psi(t)\rangle|^2  \sim t^{-0.75}$,
the return probabilities.
To our knowledge, there is no similar study of a KAM system so globally
characterized by a single fractional exponent. A mixed phase-space
regime (with a mixture of a chaotic regions and stable islands) 
can indeed exhibit fractional return probabilities \cite{Ketz2}, but it will
typically have many competing exponents, characterizing only 
local regions of phase
space. The quantum localization properties are thus not generic, and depend
sensitively on the detailed phase-space structure.

For the $2\delta$-KR, we present evidence indicating that the 
localization properties are related to the scaling properties of 
phase space in the vicinity of
golden-ratio cantori. We argue that the observed behaviour is due to
the presence of families of golden ratio cantori occupying an appreciable
fraction of each cell. 
In addition, the `cellular' momentum structure suggested by the experiments
is shown to correspond to a time-evolution operator with a 
not previously investigated 
{\em periodically-oscillating} banded structure. The corresponding
eigenstates can be well localized within a single cell defined by this
oscillating band; or they may extend over several cells.

We characterize the spectral behaviour by two parameters: a filling 
parameter, $R$, which quantifies the fraction of a cell that
a typical state occupies,
and $d$, which quantifies inter-cell transport. 
We consider two types of distinct (but related) delocalization transitions:
(a) a $0 \to 1$ cell transition as eigenstates fill one  cell
($R \simeq 1$ but $d > 2$)
(b) a $1 \to$ several cell transition, as states delocalize from a single
cell to many, occurring for $R \gg 1$ and $d \leq 2$.

As the $0 \to 1$ cell transition occurs, all states extend into fractal
cantori-filled regions bordering each cell; we identify a regime with spectral
properties (particularly spectral variances which, while not identical, may be compared
with `critical statistics'. The latter were first studied near
 the critical point of the Anderson Metal-Insulator transition (MIT)
\cite{Shapiro,Chalker,Huck,Krav,Braun,Evers}. So-called `critical statistics' are
now of much current interest
in chaotic systems with classical discontinuities (non-KAM systems) 
\cite{Bogo,Wiersig,Verbaarschot,Garcia}; they have been attributed to the effect of cantori 
\cite{Verbaarschot,Garcia}; however 
they are not expected in KAM system, due to their non-generic properties.
Below, we use the term critical statistics in this broader sense, rather than
specifically  the critical point of the MIT.

Imposing periodic boundary conditions in momentum space
effectively confines the system's eigenstates to a single cell
(with toroidal geometry). Such a calculation
shows a transition from Poisson to Wigner-Dyson statistics (they are
of similar functional form to GOE results, so are referred to as such below, 
although strictly they are COE statistics), via a new regime of 
intermediate  statistics. We also calculate eigenstates for
the open system (non-periodic boundary conditions). This calculation
yields agreement with the single-cell calculation up to the 
onset of delocalization, but beyond shows rather different behaviour.
For $d \simeq 2$ the spectral statistics show a 
signature of the onset of delocalization of the eigenstates into multiple
cells, characterized by a return towards Poissonian statistics. 

In summary, our main new results are  1) A study of the cellular form of the
unitary operator of this system, which differs substantially with
the behaviour of band random matrices (BRM), which  approximate the
ordinary QKR.
2) Numerical evidence for behaviour analogous to
critical statistics associated with cantori \cite{Garcia}.
3) The spectral signatures of delocalization from one cell, to several cells. 
4) A novel regime of exponential localization
determined by the fractional exponent $0.75$, which coincides closely with
the dominant scaling exponent obtained in the vicinity
of  golden-ratio cantori.
In Section II, we review the present state of knowledge 
of the 2$\delta$-KR. In Section III we introduce the time-evolution
operator and obtain, for the first time, an analytical form for its 
bandwidth. In Section IV we investigate the dynamical
localization and the fractional exponent $\nu =0.75$.
In Section V we look at the delocalization within a
single cell, by solving the problem `on a torus' in momentum space
and compare with critical statistics.
In Section VI we compare the behaviour with a 
calculation with `open' boundary conditions and the
signature for the onset of delocalization onto many cells.
Finally in Section VII we give our conclusions and discussion.
 
\section{The 2$\delta$-KR}

We consider a system with a Hamiltonian corresponding to a sequence of closely
spaced pairs of kicks:
\begin{eqnarray}
H= \frac{p^2}{2} - K \cos x 
\sum_n \left[ \delta(t-nT)+ \delta(t-nT+\epsilon) \right]\nonumber , 
\end{eqnarray}
where $\epsilon \ll T$ is a short time interval and $K$ is the kick-strength.
We now use a re-scaled time in units of $T$.
The classical map for the 2$\delta$-KR is then a two-kick map:
\begin{eqnarray}
p_{j+1} &=& p_j + K\sin x_{j}  \nonumber \\
x_{j+1} &=& x_j + p_{j+1}\epsilon \nonumber \\
p_{j+2} &=& p_{j+1} + K\sin x_{j+1}\nonumber \\
x_{j+2} &=& x_{j+1} + p_{j+2}(1-\epsilon) \nonumber .\\
\label{eq1}
\end{eqnarray}

Clearly, the limit $\epsilon=1$ or $0$  corresponds to the Standard Map,
which describes the classical dynamics of the QKR: 
\begin{eqnarray}
p_{i+1} &=& p_{i} + K \sin x_{i} \nonumber \\ 
x_{i+1} &=& x_i + p_{i+1} 
\label{eq2}
\end{eqnarray}

An experimental realization of this system was obtained in \cite{Jones}, 
with cold 
cesium atoms in pulsed standing waves of light. Atoms with momenta
$\langle p \rangle = p_0 \simeq (2m+1)\pi/\epsilon$ 
(relative to the optical lattice)
are confined in momentum trapping regions and absorb little energy;
conversely, atoms prepared near
$p_0\epsilon = 2m\pi$ experience rapid energy growth up to localization.
The experimental trapping regions are shown in Fig.\ref{Fig1}.
The basic mechanism of trapping is fairly intuitive: atoms for which 
$p_0 \epsilon = (2m+1)\pi$ and $m=0,1, \dots$ experience an 
impulse $K\sin x$ followed by another one $\simeq K\sin(x + \pi)$
which in effect cancels the first.
Over time, however, there is a gradual de-phasing 
of this classical `anti-resonant' process.
A theoretical study of the classical diffusion over longer times than
a couple of kicks in \cite{Jones} found anomalous momentum
diffusion for all $p_0$, with long-ranged corrections to the 
$D \simeq K^2/2$ uncorrelated diffusion rate, not present
in the Standard Map. 

Below we present the corresponding quantum behaviour.

\section{TIME-EVOLUTION OPERATOR for 2$\delta$-QKR}

The time evolution operator for this system may be written:
\begin{eqnarray}\label{eq31}
\hat{U}^{\epsilon} = \exp{-i \frac{\hat{p}^2 (T-\epsilon)}{2\hbar}} 
 \exp{i\frac{K}{\hbar} \cos x} \\ \nonumber    
\exp{-i \frac{\hat{p^2} \epsilon}{2\hbar}} \ \exp{i\frac{K}{\hbar} \cos x} \ 
\end{eqnarray}

\begin{figure}[htb]
\includegraphics*[width=0.15\textwidth,angle=-90,clip=true]{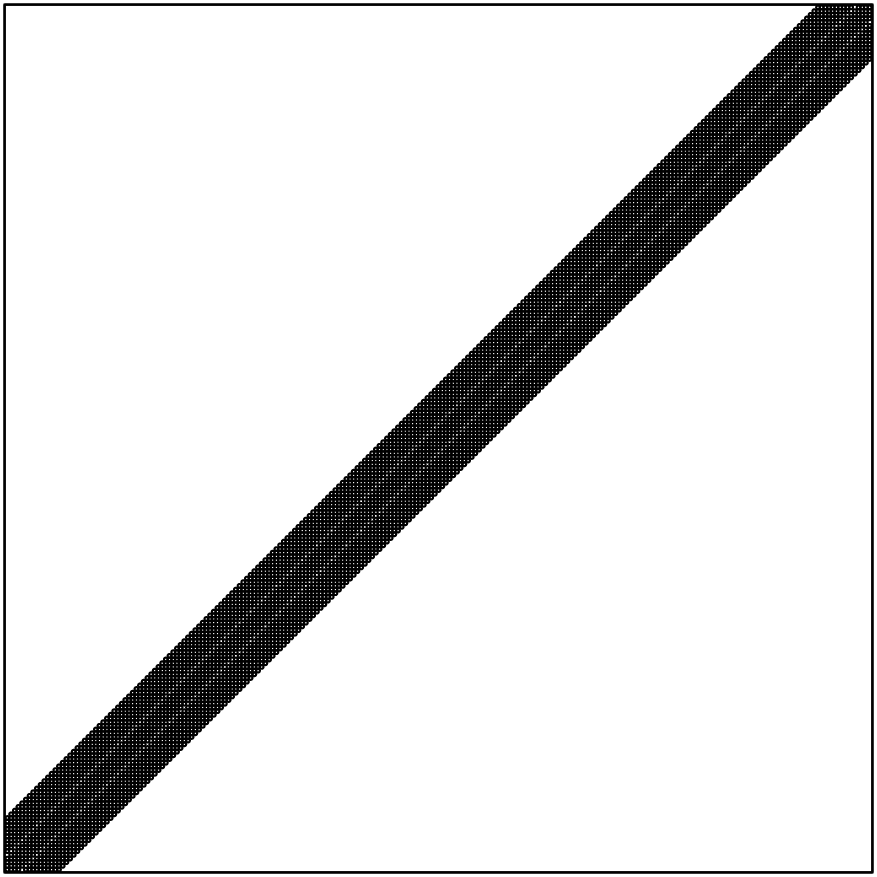}
\hspace*{5 mm}
\includegraphics[width=0.15\textwidth,angle=-90,clip=true]{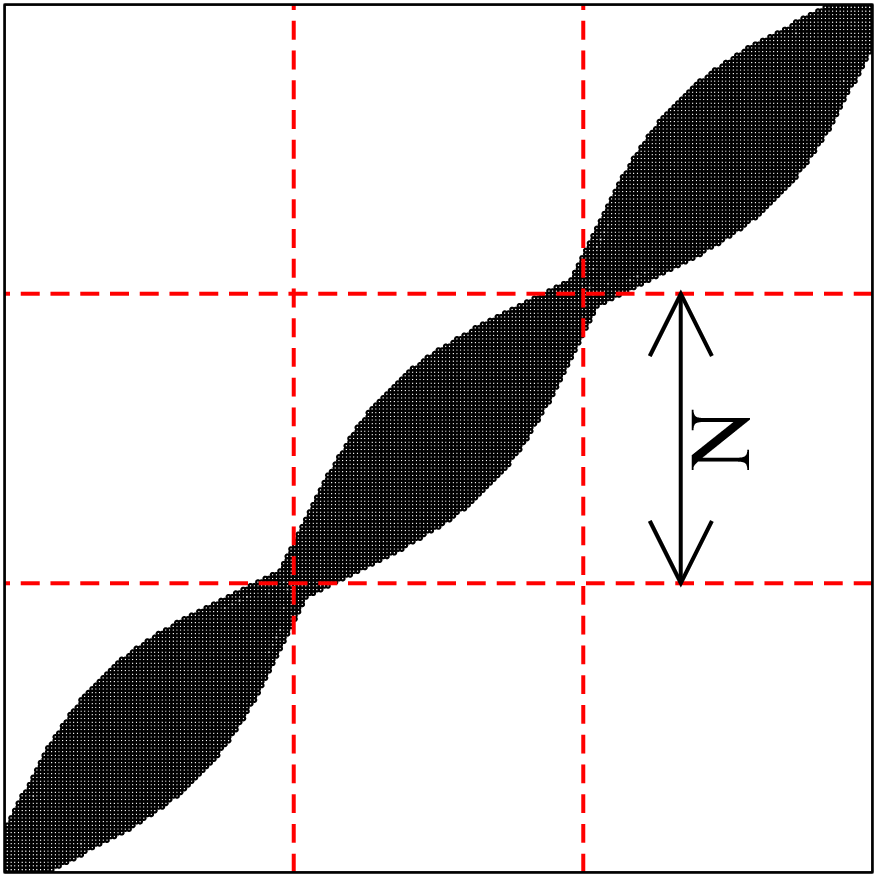}
\caption{Left {\bf(a)}: Structure of time-evolution matrix for a Quantum Kicked Rotor (QKR),
in a basis of momentum states, showing the constant bandwidth
 structure typical of a Band Random Matrix (BRM). Right{\bf(b)}: 
Form of $U^{\epsilon}$ for our system, the 2$\delta$-QKR, showing 
the oscillating bandwidth structure. Before 
delocalization, eigenstates are confined within a single `momentum cell' 
of dimension $N$.}
\label{Fig2}
\end{figure}

In a basis of plane waves,  $\hat{U}^{\epsilon}$ has matrix elements:
\begin{eqnarray}\label{eq32}
 U_{lm}^{\epsilon}=U_l^{free}. \ U_{lm}^{2-kick} = 
\exp{-i \frac{l^2\hbar (T-\epsilon)}{2}} \ i^{l-m} \\ \nonumber
 \sum_k J_{l-k}\left(\frac{K}{\hbar}\right) \ 
J_{k-m}\left(\frac{K}{\hbar}\right)
 \ e^{-i \frac{k^2\hbar \epsilon}{2}} 
\end{eqnarray}
where the $J_n(\frac{K}{\hbar})$ are integer Bessel functions
of the first kind.
It is easy to see that $U_{lm}^{2-kick}$ is
invariant if the products $K_\epsilon= K\epsilon$ and 
$\hbar_\epsilon= \hbar\epsilon$
are kept constant; while the free propagator
$U_l^{free}=e^{-i \frac{l^2\hbar (T-\epsilon)}{2}}$  
simply contributes a near-random phase. Provided that 
$l^2T \hbar \gg 2\pi$,
the results are quite insensitive to the magnitude of $(T-\epsilon)\hbar$.
Hence we often find it useful to 
consider the two scaled parameters $K_\epsilon$ and $\hbar_{\epsilon}$, 
rather than to vary $K,\epsilon$ and $\hbar$ independently.

The result in Eq.(\ref{eq32}) may be compared with the 
one-kick map in Eq.\ref{eq2}
\begin{eqnarray}
 U^{(0)}_{lm}= \exp {-i \frac{l^2 T \hbar}{2}} \ 
J_{l-m}\left(\frac{K}{\hbar}\right) .
\label{eq33}
\end{eqnarray}
The one-kick matrix for the QKR has a well-studied band-structure: since $J_{l-m}(x)\simeq 0$
for $|l-m| \gg x$, we can define a bandwith for $U^{(0)}$, namely,
 $b= \frac{K}{\hbar}$ 
(this is strictly a {\em half}-bandwidth) which is independent
of the angular momenta $l$ and $m$. However, this is {\em not} the case for 
the matrix of $U^{\epsilon}$.

It is shown in the Appendix that, assuming $|l-m|$ is small, we 
can write:
\begin{equation}
U_{lm}^{\epsilon} \approx  e^{-i \Phi}  \  J_{l-m}\left( \frac{2K}{\hbar}
\cos \left [{l\hbar \epsilon/2}\right] \right)
\label{eq34}
\end{equation}
where the phase 
$\Phi={\frac{\hbar}{2}[l^2T+ \epsilon l m + \epsilon l^2]+ \pi(l-m)}/2$.
Hence we infer a momentum dependent bandwidth, 
$b(p)= \frac{2K}{\hbar}\cos{p \epsilon/2}$. 
Fig.\ref{Fig2} shows the calculated form of both matrices 
(white denotes matrix elements less than a small threshold).
While $U^{(0)}$
has a constant bandwidth, the bandwith for the matrix of 
$U^{\epsilon}$  oscillates with $l$
from a maximum value $b_{max}=\frac{2K}{\hbar}$, equivalent to 
twice the bandwidth of $U^{(0)}$, down 
to a minimum value $b_{min} \sim 0$.

In effect, since $b_{min} \sim 0$, $U^{\epsilon}$ is partitioned into sub-matrices of dimension
$N =\frac{2\pi}{\epsilon \hbar}$ corresponding precisely to the momentum
cells of width $\Delta p = N \hbar$ observed in the experiment. 

For the QKR, the localization properties of the eigenstates
have been investigated extensively (see e.g. \cite{Izraelev} for a review).
The eigenstates are exponentially localized, with momentum probability 
distributions
$N(p) \sim \exp -2|p|/L$, where the localization length 
$L \approx \frac{K^2}{4\hbar}= b^2\hbar/4 $, in the large $K$, small $\hbar$
limits.

For the 2$\delta$-QKR, in the limit of small 
bandwidth $\frac{2K}{\hbar} \ll N$ we define a `local' localization
length for the eigenstates:
\begin{equation}
L(p) = b^2(p)\hbar/4= \frac{K^2}{\hbar} \cos^2{p \epsilon/2}
\label{eq8}
\end{equation}

This corresponds quite well with the oscillations seen in the experiment
in Fig.\ref{Fig1}(a); the energy oscillates sinusoidally 
from a maximum value $\langle p^2 \rangle \simeq 400 \sim 4L^2 \simeq 4K^4$
for $K=3.2$, $\hbar=1$, (in re-scaled units)
down to a minimum value $\langle p^2 \rangle \simeq 0$ for
$P_0 \simeq \pi/\epsilon$. In contrast, Fig.\ref{Fig1}(b)
corresponds to a regime where
the eigenstates are tending to fill each cell, i.e. $L(p) \to N\hbar$.

In \cite{Feingold,Izraelev} it was shown that the eigenstates of the QKR can
also be obtained using a $U^{(0)}$ matrix of {\em finite} dimension
$N$. The ratio of the localization length to $N$, 
was used to characterize the degree of `filling' of the matrix $U^{(0)}$
It was shown that for a ratio $R \ll 1$ the behaviour is Poissonian. With
increasing $R$, a transition to GOE behaviour was observed

For the 2$\delta$-QKR we can also introduce a  
`filling factor' $R$ defined by the ratio \cite{Feingold}:
\begin{equation}
R_{\epsilon}= \frac{K^2}{N \hbar^2} = 
\frac{K^2_{\epsilon}}{2 \pi \hbar_{\epsilon}}.
\end{equation}

While for the 2$\delta$-QKR we have a natural choice of $N$, 
determined by the position of physical boundaries (the trapping regions),
the value of $N$ in the QKR case is quite arbitrary.
The problem is solved `on a torus' in momentum.
The momentum periodicity of the matrix is adjusted by a choice of
a rational value of the kicking period $T\hbar$ (see Sec. V below).
We can then compare matrices for both $U^{(0)}$ and $U^{\epsilon}$,
with similar $N$ and $R$. 

The interesting aspect of the 2$\delta$-QKR  system is that, as we show below,
 we can vary the coupling between the cells 
 independently (with some constraints on allowable parameters)
from the degree of filling of each individual cell.
We will show below, that there is a particularly interesting regime where, for
most eigenstates, $L(p) \sim N\hbar$, but the states are still
largely trapped within a single cell. We can then `open' the boundaries
of the cells and investigate the delocalization regime.

In order to investigate the transport between cells, we examine
the surprisingly different (relative to the QKR) process of dynamical
localization in the 2$\delta$-QKR.

\section{DYNAMICAL LOCALIZATION}

\begin{figure}[htb]
\includegraphics[height=3in,width=3.in]{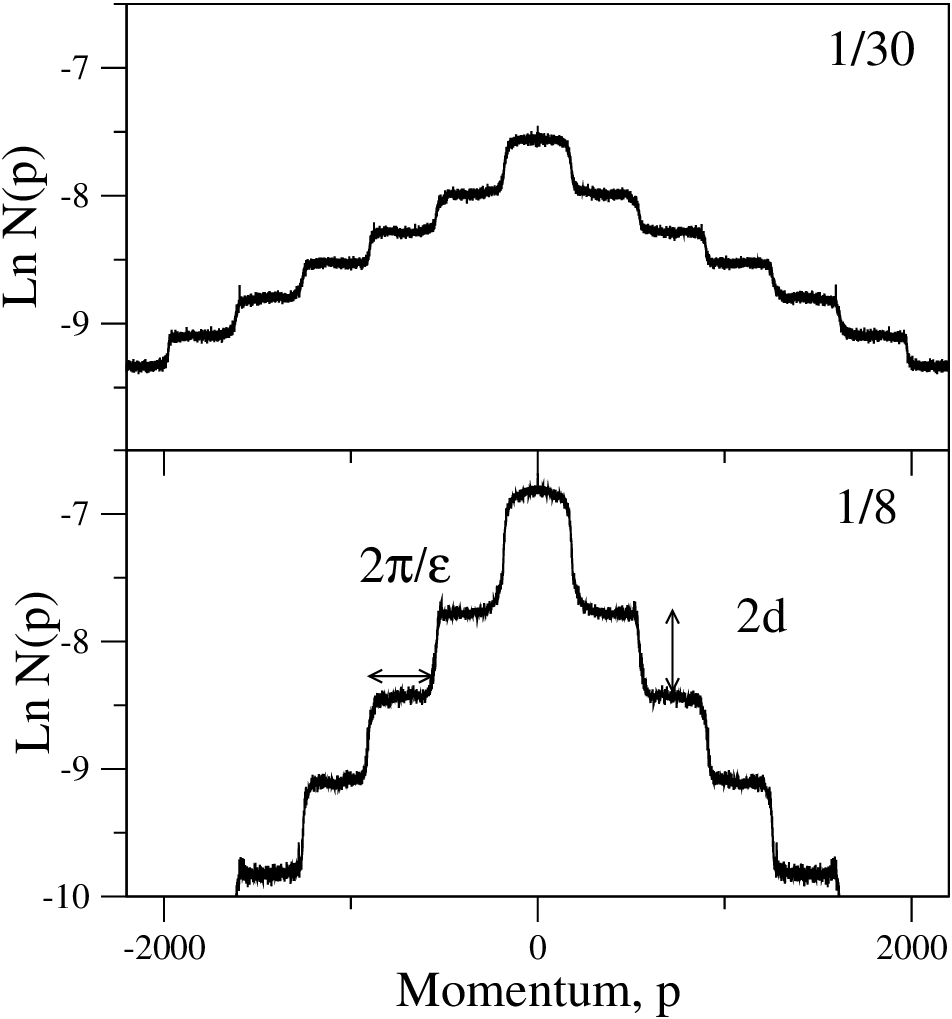}
\caption {Typical  momentum distributions,$N(p)$, 
found after dynamical localization, 
for quantum wavepackets of the 2$\delta$-QKR.
Here $K=20$, $\epsilon=0.0175$ and $\hbar=1/8$ and $ 1/30$ respectively.
The $N(p)$ of the 2$\delta$-QKR (slightly smoothed;
for both eigenstates and wavepackets), show a long-range
tail of `staircase' form which on average follows the exponential
$N(p) \sim \exp{-2(p-{\overline{p}})/L_p}$ where 
$L_p/2= N\hbar/2d$; since $N\hbar=2\pi/\epsilon$, 
the $\hbar$-dependence of $L_p$ is  determined
by the drop in probability, $d$, at each cell boundary.}
\label{Fig3}
\end{figure}
\begin{figure}[htb]
\includegraphics[width=3.5in]{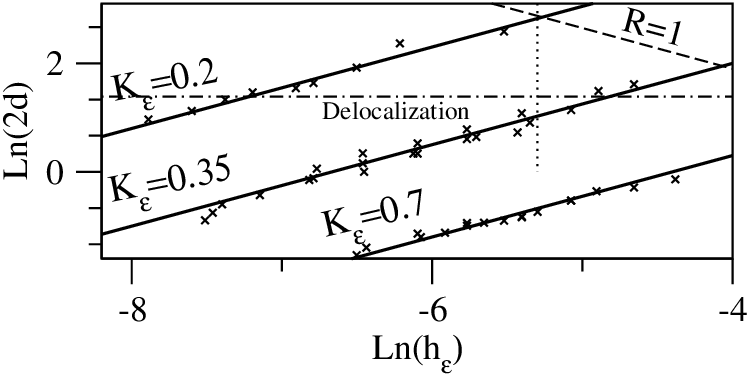}
\caption {Figure shows that $Ln(2d)$ plotted against $Ln (\hbar \epsilon)$ 
lies close to straight lines of invariant $K_{\epsilon} = K \epsilon$, 
with constant slope $\approx 0.75$. Hence $d \propto (\hbar \epsilon)^{0.75}$
and $L_p \propto \hbar^{-0.75}$ -in contrast to the well-known
QKR result $L_p \propto \hbar^{-1}$.
The dashed line indicates $R \simeq 1$  the one-cell filling border 
(i.e. $R > 1$ below the line). 
The {\em delocalization} border $(d \simeq 2)$ is the dot-dash line
(i.e. $d<2$ below the line
and represents the onset of significant coupling between cells).
Statistics are presented later, in Fig.\ref{Fig13},
for points corresponding to the dotted line. }
\label{Fig4}
\end{figure}

A set
of wavepackets (all initially with $N(p, t=0) \simeq \delta(p)$)
were evolved
in time, using the time evolution matrix $U^{\epsilon}$, for a range of $K$,
$\epsilon$ and $\hbar$. Fig.(\ref{Fig3}) shows typical momentum
distributions obtained after a long time (beyond the `break-time' $t^*$
for the onset of dynamical localization).
Fig.\ref{Fig3} shows some typical momentum distributions $N(p)$.
They are modulated by an exponential envelope
\begin{equation}
N(p) \sim \exp{-\frac{2(p-{\overline{p}})}{L_p}}
\label{eq10}
\end{equation}
but with a regular `staircase' structure superposed. There is a 
steep drop in probability at each step:
\begin{equation}
N^+(p) = \exp{-2d} \  N^-(p) ,
\label{eq11}
\end{equation}
where $N^\pm (p)$ denote the probability after and before the step
respectively. The localization length is
\begin{equation}
L_p =\frac{ 2\pi}{d\epsilon} .
\label{eq12}
\end{equation}
The parameter $d$ controls the transport through
the cell boundaries. It also contains the $\hbar$ dependence
of $L_p$.
In Fig.\ref{Fig4}, the dependence of $d$ on $K_{\epsilon}$
and $h_{\epsilon}= \hbar \epsilon$ is shown. 
It may be seen that to very good accuracy:
\begin{equation}
d \propto \frac{( \hbar \epsilon)^{0.75}}{f(K_{\epsilon})}
\end{equation}
where $f(K_{\epsilon})$ is some function of $K \epsilon$. 
A rough fit yields an estimate, to within $50\%$ or so:
\begin{equation}
d \simeq \frac{3.5 ( \hbar \epsilon)^{0.75}}{K_{\epsilon}^3} .
\label{eq12b}
\end{equation}

Hence we obtain the surprising result that for the 2$\delta$-QKR, 
the localization length has an $\hbar$ dependence with a {\em fractional} power,
$L_p \sim \hbar^{-0.75}$. In comparison, for the QKR, $L_p \sim \hbar^{-1}$.

It is interesting to consider the origin of the $\hbar^{-0.75}$ behaviour.
In fact, the first $\sim 1-2$ steps of the staircase can be seen in experiments
with optical lattices of \cite{Jones}.
An earlier experiment \cite{Christensen} with pairs of broad pulses 
(as opposed to pulses short enough in duration to approximate 
$\delta$-kicks) also showed a
steep drop in probability over a narrow momentum region. This was identified as
due to the presence of cantori. The finite pulse system does not have the
momentum periodicity (and hence the generic character)
of the $2\delta$- kicked system,
and  is effectively integrable at large $p$. Nevertheless, the
classical dynamics is similar to the $2\delta$-QKR for $p \sim 0$. 
The QKR shows analogous behaviour: for pulses of finite duration, the
classical dynamics is similar to the $\delta$-kicked system for $p$ small.

We can examine the classical behaviour in the trapping regions.
Starting from the two-kick map Eq.\ref{eq1} and expanding
our initial momenta  around the trapping
momenta $p_j \simeq P_n +\delta p^j$ where $P_n =(2n+1)\pi/\epsilon$, 
we can write
\begin{eqnarray}
p_{j+2} = p_{j} + K\sin x_{j} + \nonumber \\
         K\sin \left[ x_j + (2n+1)\pi +\delta p^j \epsilon + 
          K \epsilon \sin x_{j} \right] .
\end{eqnarray}

We can expand the trigonometric expressions, making
small angle approximations if we assume
$\delta p^j \epsilon \ll 2\pi$ and $K \epsilon \ll 2\pi$.
In the trapping regions we then obtain an approximate one-kick map
\begin{eqnarray}
p_{j+2} &\simeq& p_{j} - K^2 \frac{\epsilon}{2} \ \sin 2 x_{j} - K \delta p^j \  \epsilon \cos x_{j}\label{eq13} \\
x_{j+2} &\simeq& x_{j} + p_{j+2} T 
\label{eq13a}
\end{eqnarray}

The small angle assumption $\delta p^j \epsilon \ll \pi$ 
constrains $\Delta p$, the effective width
of the trapping region,
\begin{equation} 
\Delta p  \sim 1/ \epsilon .
\label{eq14} 
\end{equation}

\begin{figure}[htb]
\includegraphics*[height=1.5in]{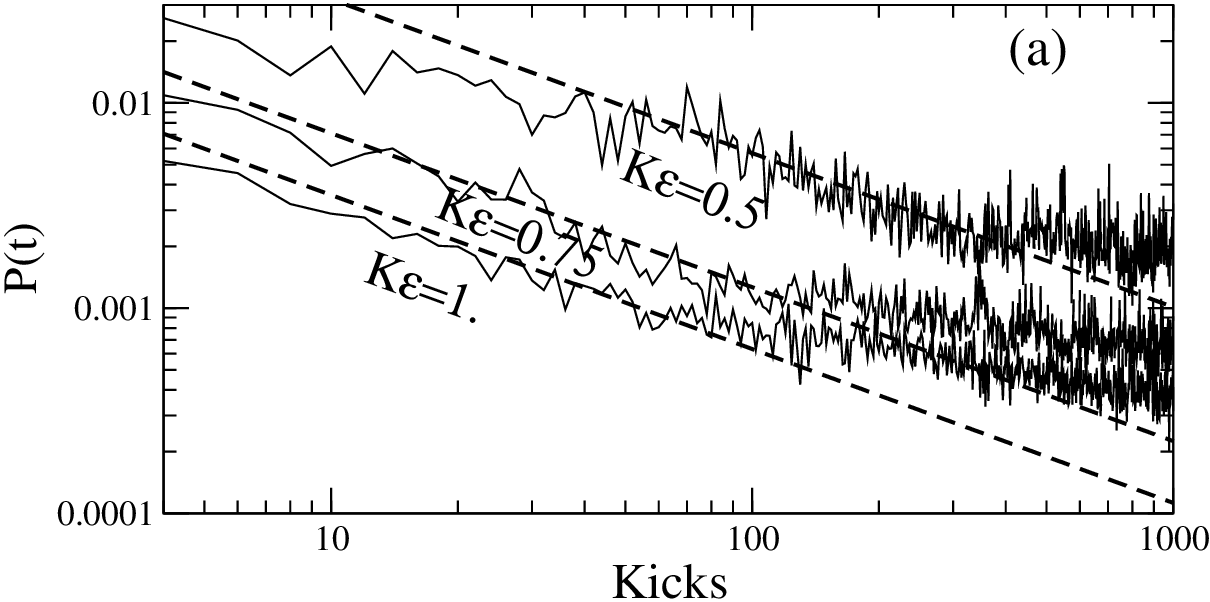}
\vspace*{5 mm}
\includegraphics[height=1.5in]{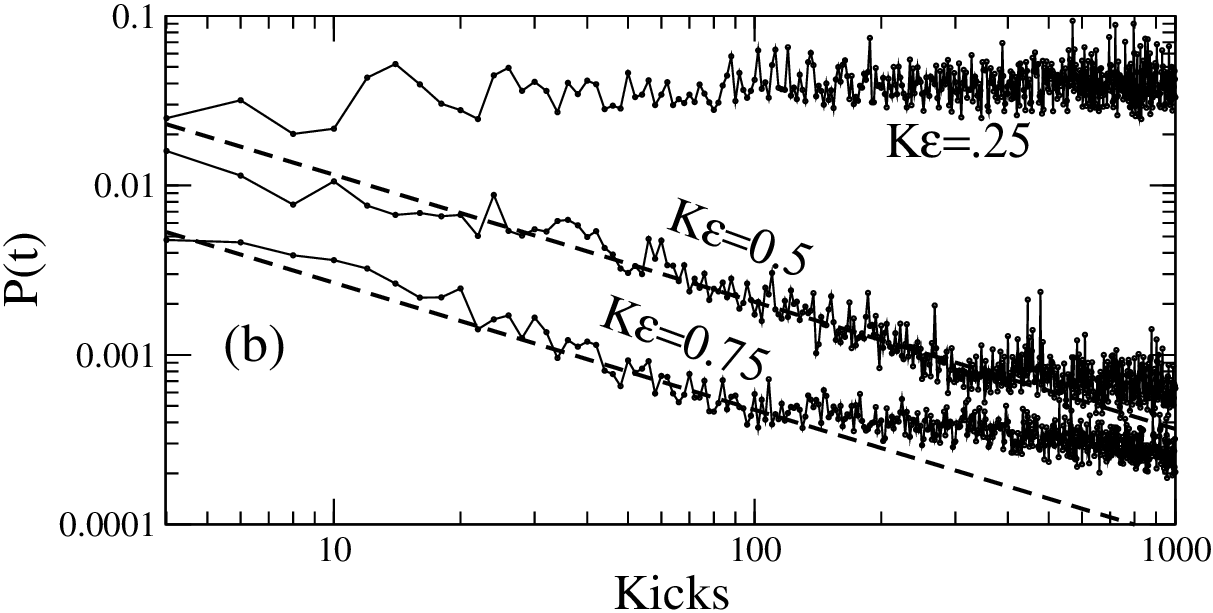}
\caption{Return probabilities for wavepackets prepared near
the centre of the trapping
regions indicating a $P(t) \sim t^{-0.75}$ decay, before  localization.
The dashed straight lines indicate a slope $=0.75$.
Note that at the lowest value of $K\epsilon=0.25$,
wavepackets prepared at the centre of the trapping region 
do not spread in momentum.
{\bf(a)} $\hbar=1/8$ {\bf(b)} $\hbar=1/16$}
\label{Fig5}
\end{figure}

A study of the Poincar\'e surfaces of section (SOS) may be 
found in \cite{Mischa}.
But in summary, at the centre of the trapping region, where 
$\delta p^j=0$, only the $\sin 2x$ kick in Eq.\ref{eq13} is significant
so the classical SOS shows structure very
similar to a `period-doubled' Standard Map with impulse
$V'(x)=K_{eff} \sin 2 x $ where
$K_{eff} \simeq K^2 \frac{\epsilon}{2}$
and $K_{eff}$ is an effective kick strength.
However, further out within the trapping region,
the SOS show that it is the $\cos x$ kick in Eq.\ref{eq13} which is dominant
and the island structure
is similar to that of a Standard Map with impulse:
\begin{equation}
V' (x)\simeq K \delta p \  \epsilon \cos x.
\label{eq15}
\end{equation}

Given the importance of the $K \epsilon$ scaling seen in Fig.\ref{Fig4}
(rather than a $K^2\epsilon$ scaling) we suggest
 that the $K_{eff} \cos x$ form of Eq.\ref{eq13}
determines the quantum behaviour. This implies that regions with:
\begin{equation}
K \delta p^j \  \epsilon  \gg K^2 \frac{\epsilon}{2} 
\end{equation}
dominate transport.
This will be the case if $\delta p_j$ is large; from Eq.\ref{eq14},
we deduce that this implies the criterion $K \epsilon \ll 1$. 
However, if $K \epsilon$ is too small, the phase space will be too regular.
Hence, values of $K \epsilon \approx 0.1 - 0.3$ seem indicated.

In any case, the resonance structure in much of the trapping regions is always 
locally similar to the Standard Map, but with a varying effective kick strength
(and phase of the impulse). For the Standard Map the last invariant phase-space manifolds correspond to $p/(2\pi) \simeq \phi$ or 
$\simeq 1-\phi$ where $\phi$ is the Golden Ratio.
The fractal remnant of this last manifold plays a role in transport
in the Standard Map for $K \simeq 1-3$. Here, this would suggest 
$\delta p K \epsilon \approx 1-3 $ and $\delta p \sim 10$.

The scaling properties of the phase-space
around the golden ratio cantori 
were investigated by \cite{Shmuel87}. A characteristic exponent
$\sigma_s \approx 0.75$ was found in directions dominated by elliptic
fixed points. We term this the stable/dominant exponent.
On the other hand $\sigma_u \approx .66$ was found in directions
dominated by hyperbolic fixed points. We term this the 
unstable/sub-dominant exponent.

Previous studies of the QKR \cite{Geisel} 
found a $L \propto \hbar^{+\sigma}$ scaling,
with $\sigma \simeq \sigma_u \simeq 0.66$,
for momenta near the range $ 1-\phi < pT/2\pi < \phi$.
In \cite{Maitra}, a $L \propto \hbar^{+\sigma_u}$ scaling
was associated to a tunneling type process (termed `retunneling')
in \cite{Maitra}. An abrupt change to
a regime with $L \propto \hbar^{-\sigma}$, with $\sigma \approx 0.5$,
was also observed,
and was attributed to a localization process which dominates
when transport through the fractal cantoral regions is more `open'.
In that work, it was argued that the process has similarities with
dynamical localization.

The $2\delta$-KR results, with a negative sign on the exponent 
$L \sim \hbar^{-0.75}$ correspond
to a localization regime, and are consistent with quantum
probability `sticking' mostly to directions where there are (or were,
at lower $K$) elliptic fixed points. One might speculate why no previous
studies uncovered a regime dominated by $\sigma_s$; 
we note that in \cite{Geisel,Maitra}
the $L \sim \hbar^{+0.66}$ scaling refers to a local region; these
are regimes where there are many stable islands and overall diffusion
in each phase-space manifold is either absent, or rather slow.
Here, in contrast, part of phase-space is taken by very chaotic,
fast diffusing regions, with trajectories which roam freely over large areas
of phase-space; the only appreciable localization occurs at the
stabler parts of cantoral remnants. 
In many regimes, there can be contributions from $\sim 1/\epsilon$ groups
of golden-ratio cantori, rather than simply a single
region near $p/2\pi  \simeq \phi$ like the Standard Map. 
We note that Fig.13 in \cite{Maitra}, where
the $\sigma=0.5$ scaling is obtained, is in a difficult numerical
regime: barring one data point, a value $\sigma \simeq 0.75$ is not
implausible. 

We also calculated return probabilities $P(t)$ of the quantum wavepackets 
with time,
averaged over 100 initial starting conditions close to the centre of the 
trapping region, 
\begin{equation}
P(t)=  \left( 1/100 \right) \sum_{l_1}^{l_2} \langle P_l(t) \rangle ,
\label{eq14a}
\end{equation}
where $P_l(t)= |\langle \psi(t)|\psi_l(t=0) \rangle |^2$. The initial
condition was taken to be an angular momentum eigenstate
$\psi_l(t=0) = |l \rangle$.
The results were averaged from $l_1=\pi/\epsilon\hbar$ to $l_2=l_1+100$.
Fig.\ref{Fig5} shows plots of $P(t)$, which show that
$P(t)$ decays as $t^{-0.75}$ up to the break-time.
For the lowest values of $K\epsilon$, the $t^{-0.75}$ decay is not
apparent since the wavepacket localizes almost immediately.

A well-known relation between power-law return probabilities and the
spectral statistics (variances) has been investigated for `critical statistics'
in non-KAM billiards or the 
Anderson transition \cite{Chalker,Krav, Evers,Garcia}.
While the present systems has important differences (it is 
a smooth KAM system; it has a cellular structure and an
oscillating band unitary matrix), the pre-eminence of the single
fractional exponent motivates an investigation of the spectral
properties.

We investigate the statistics as a function of the 
filling factor $R$ and the inter-cell transport parameter $d$.
Calculations were done for two types of boundary conditions:
periodic boundary conditions and `open' (non-periodic) boundary 
conditions. These are discussed in turn below.

\section{EIGENSTATES: PERIODIC BOUNDARY CONDITIONS}

\begin{figure}[h]
\includegraphics[width=3.5in]{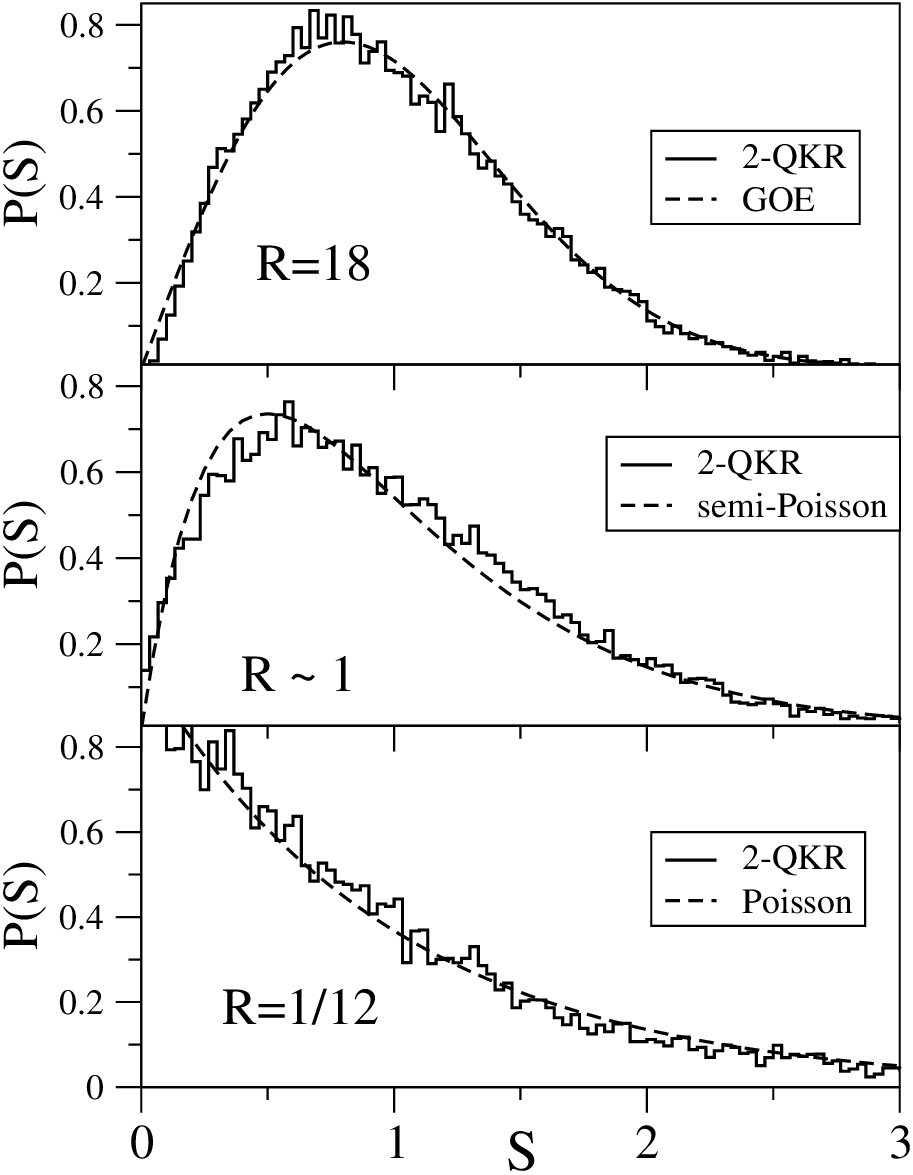}
\caption {Typical 
nearest-neighbour distributions $P(s)$, for periodic boundary
conditions, for the $2\delta-KR$ 
as a function of the filling factor $R$. The distributions are
Poissonian for small $R$, intermediate for $R \sim 1$ and GOE
for sufficiently large $R$. Although the intermediate case  never
exactly follows the semi-Poisson distribution
$(P(s)=4s e^{-2s})$, a comparison is useful, and so is shown
in the middle figure}
\label{Fig6}
\end{figure}

\begin{figure}[h]
\includegraphics[width=3.in]{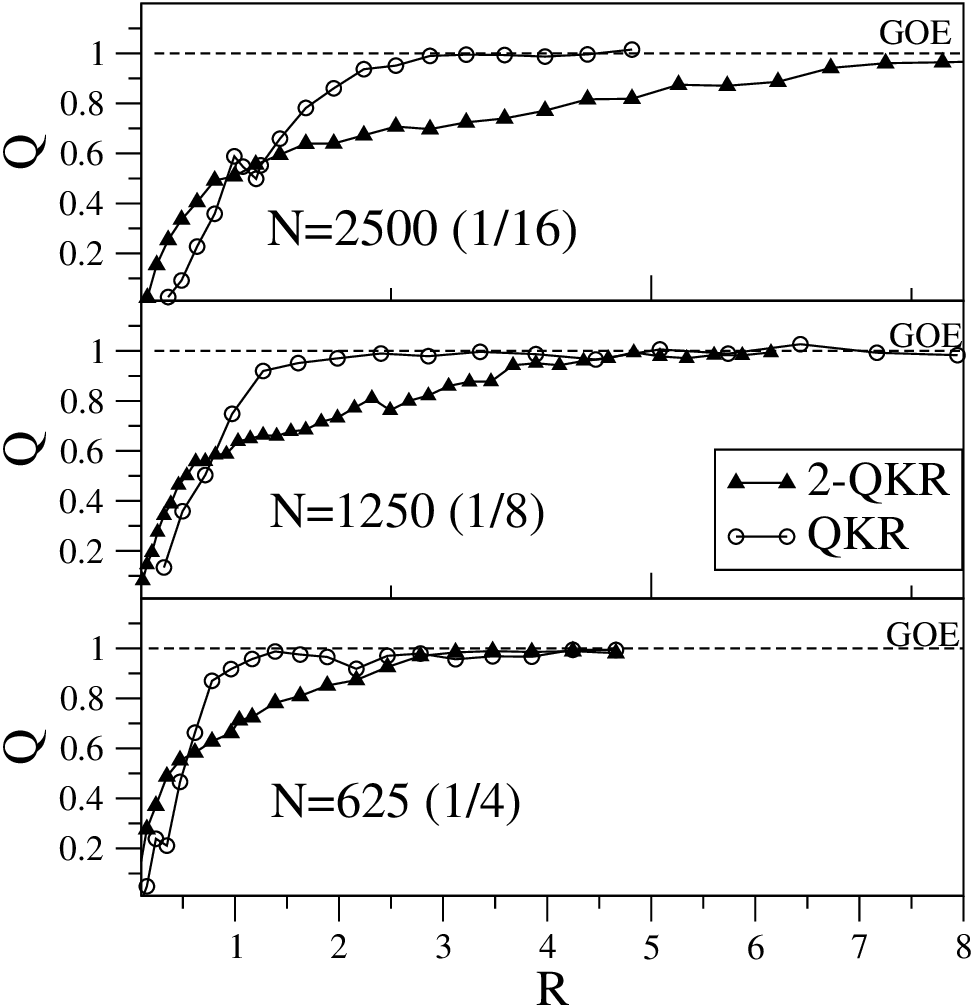}
\caption {Comparison of the transition from Poissonian ($Q=0$)
to GOE ($Q=1$) NNS statistics for the QKR relative to the 
$2\delta-KR$ as a function of the filling factor
$R=K^2/N\hbar^2$. While the QKR makes a 
rapid transition from Poisson to GOE for $R \simeq 1$,
 the $2\delta-KR$  on the other hand, shows
 a clear change in slope
where delocalization of the quantum eigenstates is hindered by the fractal cell
borders. Here $M=25$; the values shown in brackets are approximate
values of $\hbar=2\pi M/N$; to avoid
commensurability of $M$ and $N$,
actual values of $N$ used were $629, 1259 $ and $2513$. }
\label{Fig7}
\end{figure}

We may keep the dimension of the unitary matrix $U$ to a finite value $N$ 
by making the momentum periodic with period $N\hbar$,
following the approach in \cite{Feingold,Chang,Izraelev}. 
In order to preserve unitarity we must use a resonant 
value of $\hbar \epsilon$. We take:
\begin{equation}
\hbar \epsilon = \frac{2\pi}{N}
\end{equation}
and then
\begin{equation}
\hbar \tau =\hbar(T-\epsilon)= M \hbar \epsilon 
\label{eq15b}
\end{equation}
where $M$ is the closest integer to $(T-\epsilon)/\epsilon$.
We then construct a unitary matrix with elements:
\begin{eqnarray}
U_{ln}(\tau_j) = \frac{1}{N} \ \exp{-i \tau_j(l+ \theta_x)^2 /2} \nonumber \\
                        \sum_{k=-N_1}^{N_1}   
                   \exp{i \frac{K}{\hbar}} 
                   \cos {2\pi(k+ \theta_p)}
                   \ \exp{2\pi i(k+ \theta_p)(l-n)}\nonumber \\
\label{eq16}
\end{eqnarray}

We may have $\tau_1 =\hbar \epsilon$ or $\tau_2 = m \tau_1$.
Here $N_1=(N-1)/2$. We construct the full two-kick matrix
of dimension $N$
\begin{equation}
U_\epsilon (N)= U(\tau_1) \ U(\tau_2) ,
\end{equation}
which is then diagonalized to obtain $N$ eigenvalues and eigenstates.
We may compare the results with the QKR equivalent
\begin{equation}
U^0 (N)= U(T\hbar) .
\end{equation}
Here $\hbar=2 M \pi/N$, where $M$ is an integer (non commensurate
with $N$) which determines
the momentum width of the matrix, i.e. $N\hbar= 2 M \pi$; unlike the
$2\delta$-KR, there is no underlying physical cellular phase-space structure
to justify the choice of a particular $M,N$. 

Note the dependence of the matrix elements on two Bloch phases:
$\theta_p$ is a Bloch phase in the $p$ direction while $\theta_x$, the Bloch
phase in the $x$ direction, is the quasi-momentum. If $\theta_p \neq 0, \pi$, 
parity is not a good quantum number and we may use all eigenvalues 
in the statistics, regardless of parity. 
It is customary (e.g. \cite{Ketz2,Garcia}) 
to calculate spectra for several $\theta_p$ to improve significance. 
In practice, we found that for the
$2\delta-KR$, even for sizable $K$, there are very localized states which are 
too removed from the boundaries and generate pairs of parity-related pairs of
near-degeneracies. The parity conservation effect is more effectively
eliminated by, in addition to $\theta_p \neq 0, \pi$, also having
$\theta_x \neq 0, 0.5$. The cantori effects we investigate are not 
affected by a non-zero
quasi-momentum, and thus we average over several quasi-momenta. 

\begin{figure}[h]
\includegraphics[width=3.5in]{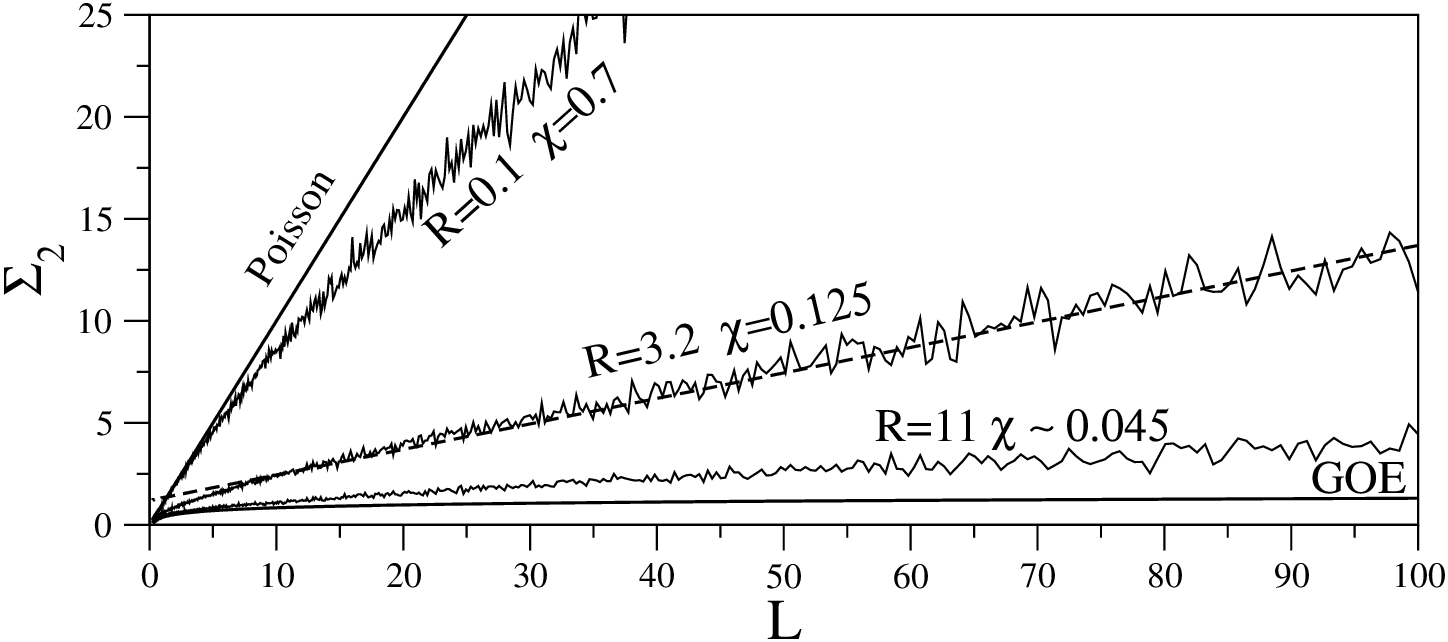}
\caption {Typical number variances $\Sigma_2(L)$
as a function of $R$ for the $2\delta-KR$.
We use the best fit to the slope to estimate the level
compressibility $\chi$. For $R$ small, we have Poissonian
statistics and $\Sigma_2(L) \simeq \chi L \simeq L$. For
sufficiently large $L$, $\Sigma_2(L) \sim Ln(L)$ is far from linear;
in this case the value $\chi < 0.05$ is simply an indication that
$\Sigma_2(L)$ is close to the GOE limit.
We identify an intermediate regime where $\Sigma_2(L) \sim \chi L$
for $1 \ll L \ll N$,(here $N=2513$) with $\chi \sim 0.13$, corresponding 
to a similar range of $R$ as the  regime seen in Fig.\ref{Fig7}.}
\label{Fig8}
\end{figure}

\begin{figure}[h]
\includegraphics[width=3.in]{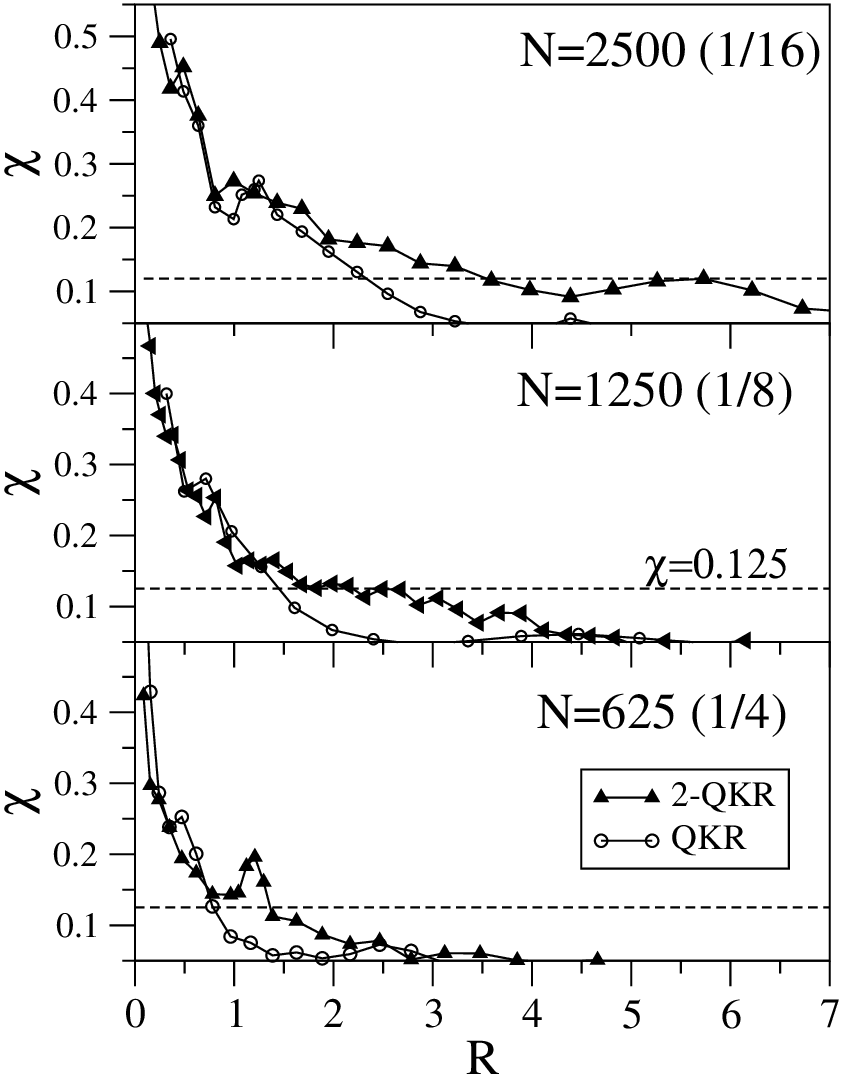}
\caption {Shows slopes ($\chi$) of the variances $\Sigma_2(L)$. Figure 
compares the transition from Poissonian ($\chi=1$)
to GOE ($\chi < 0.05$) for the QKR with those of the
$2\delta-KR$ as a function of the filling factor
$R=K^2/N\hbar^2$. Parameters are similar to Fig\ref{Fig7}. The QKR
makes a rapid transition to GOE at $R \simeq 1$ (with the
exception of the small minimum
seen at $R \simeq 1$ and $N=2500$ which corresponds to a classical
anti-resonance $K \simeq \pi$). The $2\delta-KR$ on the other hand
shows an approximate plateau near $\chi \simeq 0.125$, the value
one might expect from the `critical' statistics relation $\chi \simeq 1/2(1-D_2)$
if $D_2 \simeq 0.75$.}
\label{Fig9}
\end{figure}

\subsection{Effects of cantori in $0 \to 1$ cell transition: 
QKR versus $2\delta-KR$ }

We have calculated eigenvalues of the unitary matrix for a range of values
of $M$, $N$ and $K$. Below we compare nearest neighbour ($P(S)$) distributions
and spectral variances
$\Sigma_2(L) = \langle L^2 \rangle - \langle L \rangle^2$, 
of the $2\delta-KR$ with those of 
the QKR as a function of the filling factor $R =K^2/N\hbar^2$. 
We present results for $M=25$ (i.e. $\epsilon=0.04$)
and values {\em close} to $N=625, 1250$ and $2500$
(one cannot have $N, M$ commensurate).

\begin{figure}[htb]
\includegraphics[width=3.in]{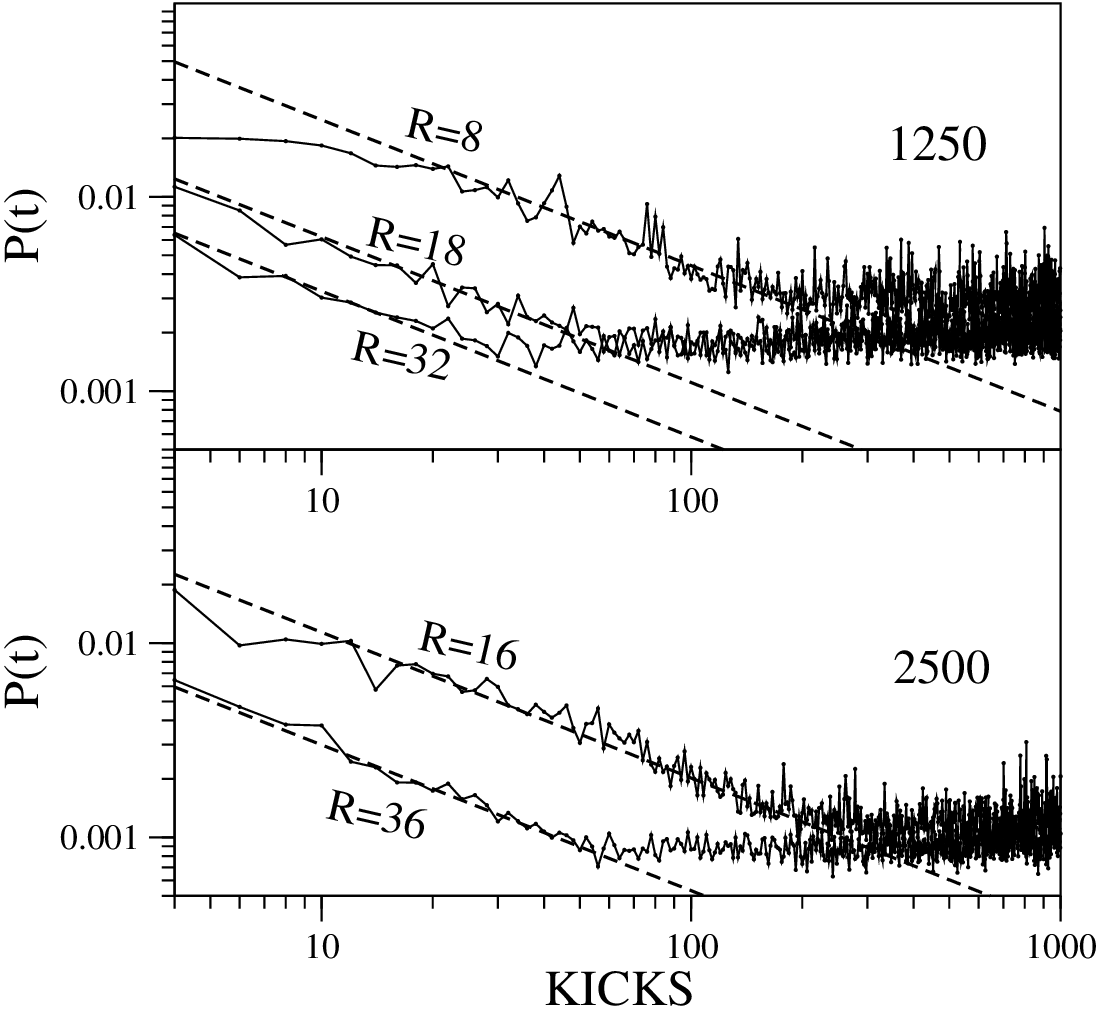}
\caption {Return probabilities for the finite dimensional 
problem, for similar parameters to Fig.5. The $P(t)$ still
decay as $t^{-0.75}$ for $t < t^*$, but $P(t) \to 2/N$ for
$t \to \infty$.}
\label{Fig10}
\end{figure}

In Fig.\ref{Fig6} we show typical NNS distributions for $R \ll 1$, 
$R \sim 1$ and $R \gg 1$. For both single/double-kicked 
systems, the behaviour evolves from
Poissonian to GOE, via an intermediate distribution at $R \sim 1$.
While the  $2\delta-KR$ does not exactly show the intermediate Semi-Poisson
form, it can get quite close, and so a comparison is helpful.
Here we quantify the deviation of $P(S)$ from 
$P_{P}(S)$ and $P_{GOE}(S)$,
its Poisson and GOE limits respectively,
with a quantity $Q$ \cite{Casati2}:
\begin{equation}
1-Q= \frac{ \int_0^{S_0} ( P(S') - P_{GOE}(S') dS')}
{\int_0^{S_0} (P_{P}(S') - P_{GOE}(S'))dS'} .
\end{equation}
Hence $Q=0$ indicates a Poisson distribution, while $Q=1$
signals a GOE distribution. We take $S_0=0.3$.
In Fig.\ref{Fig7} we plot $Q$ as a function of $R$. We find that while
the QKR moves rapidly from Poisson to GOE for $R \sim 1$, the 
$2\delta-KR$ curve abruptly changes slope and the distribution evolves
much more slowly towards GOE. We identify this as the regime where
delocalization of the eigenstates over the whole cell is constrained by 
the cantori regions bordering the cells. A fit to alternative functions, 
such as the Berry-Robnik \cite{Robnik},
gives a qualitatively
similar picture. 

\begin{figure}[ht]
\includegraphics[width=3.in]{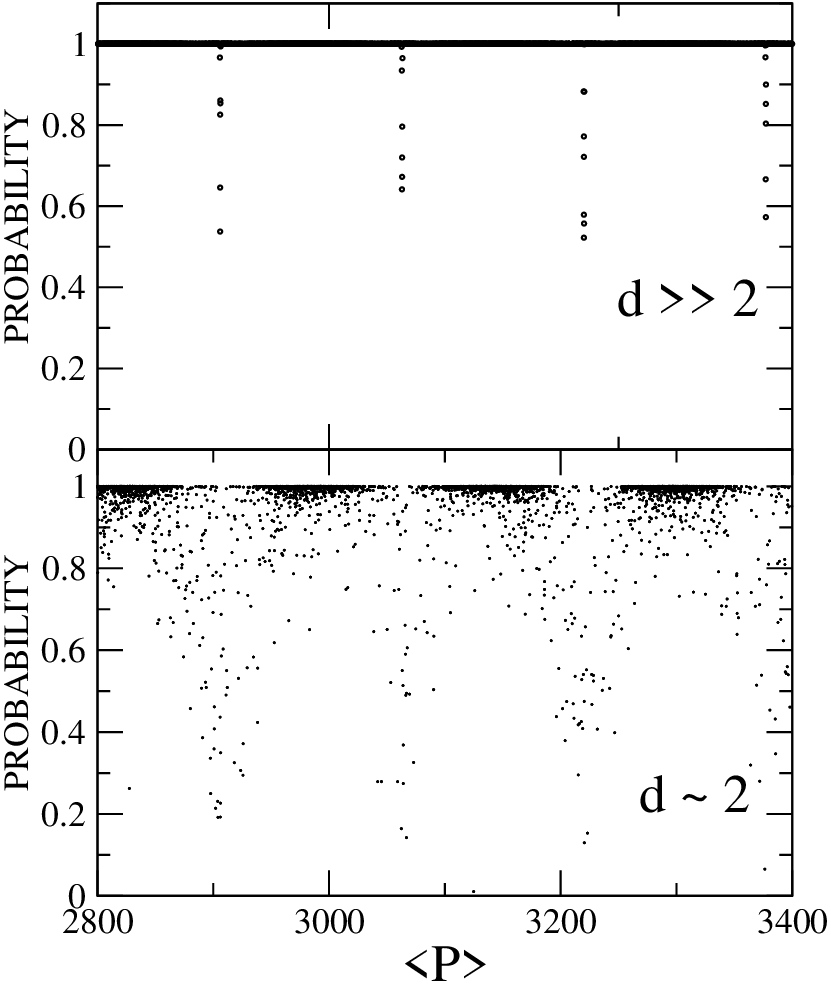}
\caption {Figure shows the effect of delocalization onto multiple cells: each point indicates
the total probability of finding a given eigenstate
in the cell which contains its average momentum $\langle p \rangle$.
Well below the delocalization border,
($d \gg 2$) all states are fully contained
in their assigned cell (barring a few edge-states
at the boundary of each cell).
But, at the `delocalization border' ($d \simeq 2$), a proportion of states
have appreciable probability in neighbouring cells -- in fact about
$2\%$ of states have less than half of their probability in their
assigned cell. Results correspond to
 $\epsilon=0.04$,$ \hbar=1/8$ and $K=3$ (upper figure)
or $K=7$ (lower figure).}
\label{Fig11}
\end{figure}

In Fig.\ref{Fig8}, we show that the variances are close to
Poissonian $\Sigma_2(L) \simeq L$ for small $R$, and are close to
GOE for large $R$. However, for $R \sim1$ there is  a regime,
with nearly linear slope,  $\Sigma_2(L) \simeq \chi L$ for $L \gg 1$
but $L \ll N$.
We fitted the slopes of the $\Sigma_2(L)$ to the best straight
line in the range $L= 7 \to 37$. We note that the graphs are not
necessarily very linear everywhere: in certain regimes
there is a pronounced curvature. Nevertheless, the procedure does give 
an indication of the average slope.
While the $N=2500$ results may remain linear out to $L \approx 100$, 
the $N=625$ results saturate at much lower $L$, hence the $L=7-37$ range
is a compromise, good for all three cell sizes.

In Fig.\ref{Fig9} we plot the slopes $\chi$ (the level compressibilities),
for the $2\delta-$KR. We see that  
above $R \simeq 1$, there is, relative to the QKR, a 
plateau in the level compressibility, where $\chi \approx 0.12$.
This behaviour is completely absent in the QKR: for $R \simeq 1$,
the values of $\chi$ evolve rapidly towards the GOE limit.
 
We recall the results \cite{Chalker} for the level
compressibilities of the Anderson MIT. It is predicted
that asymptotically the variances have a linear form, for $L \gg 1$:
\begin{equation}
\Sigma_2(L) \approx \chi L ,
\end{equation}
where 
\begin{equation}
\chi \approx 1/2(1-D_2/D) < 1
\end{equation}
and $D$ is the spatial dimension. 
$D_2$ is a multifractal exponent related to the
inverse participation ratio. This behaviour corresponds
\cite{Chalker}, in the MIT,
to return probabilities which decay as $P(t) \sim t^{-D_2}$
and is considered a `fingerprint' of critical systems, even in 
systems without disorder such as the non-KAM billiards \cite{Garcia}.

Here we calculate the return probabilities using the eigenstates 
$|\alpha \rangle$ and eigenvalues $e^{i \omega_{\alpha}}$
of the N-dimensional matrix. We calculate 
\begin{equation}
P_l(t)=  \sum_l \left| e^{i \omega_{\alpha}} 
| \langle l| \alpha \rangle |^2 \right|^2 
\label{eq15a}
\end{equation}
and average $P_l(t)$ over different starting conditions
in the trapping regions $|l \rangle$, 
exactly as in Eq.\ref{eq14a}.

We note that in Eq.\ref{eq14a}, the $P(t)$ were
obtained independently of the eigenstates by direct time evolution
of a wavepacket in the {\em unbounded} system. The behaviour
in Fig.\ref{Fig10} is very similar to that in Fig.\ref{Fig5}, and again
a decay rate $P(t) \sim t^{-0.75}$ is apparent. However, $P(t)$ in
Fig.\ref{Fig10} saturates to a larger
value than for Fig.\ref{Fig5}, where the wavepackets are not
restricted to a finite cell. For the finite $N$ case $P(t) \to  2/N$
as $t \to \infty$ (the factor of 2 is attributed to weak 
localization \cite{Smilanski}).
We also observe that for $R \simeq 1$, in the regime with $\chi \sim 0.125$,
wavepackets started in the trapping regions 
localize too rapidly to demonstrate the power-law decay.
Nevertheless the behaviour of $\chi$ is -numerically-
consistent with a fractal exponent $D_2 \simeq 0.75$.

\begin{figure}[h]
\includegraphics[width=3.5in]{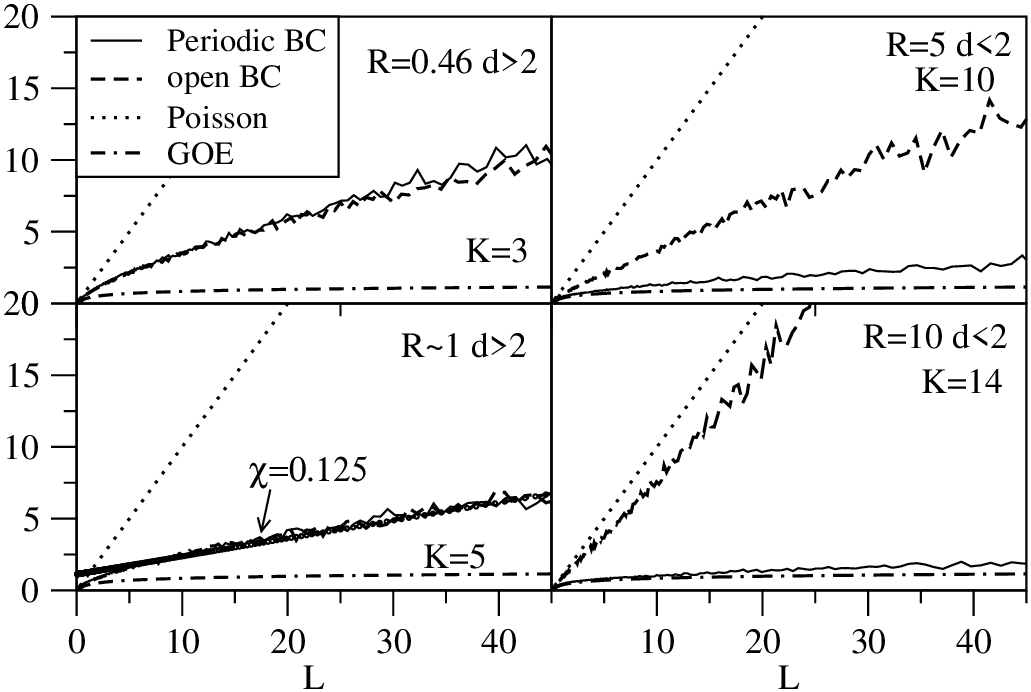}
\caption {Shows the effects of delocalization onto
multiple cells on the spectral variances. Variances obtained
with periodic boundary conditions ($N \approx 1250$) are compared with `open'
boundary conditions ($N_{tot}=10,000$ , $N \approx 1250$,$M=51$). 
For $d >2$, the eigenstates
are well confined within a single cell and there is no
significant difference between  `open' and `closed' BCs.
When $d < 2$, the results are  sensitive to the BCs.
While the `closed' (periodic) BCs results tend to GOE,
the `open' results revert towards Poissonian statistics.}
\label{Fig12}
\end{figure}

\begin{figure}[htb]
\includegraphics[width=3.in]{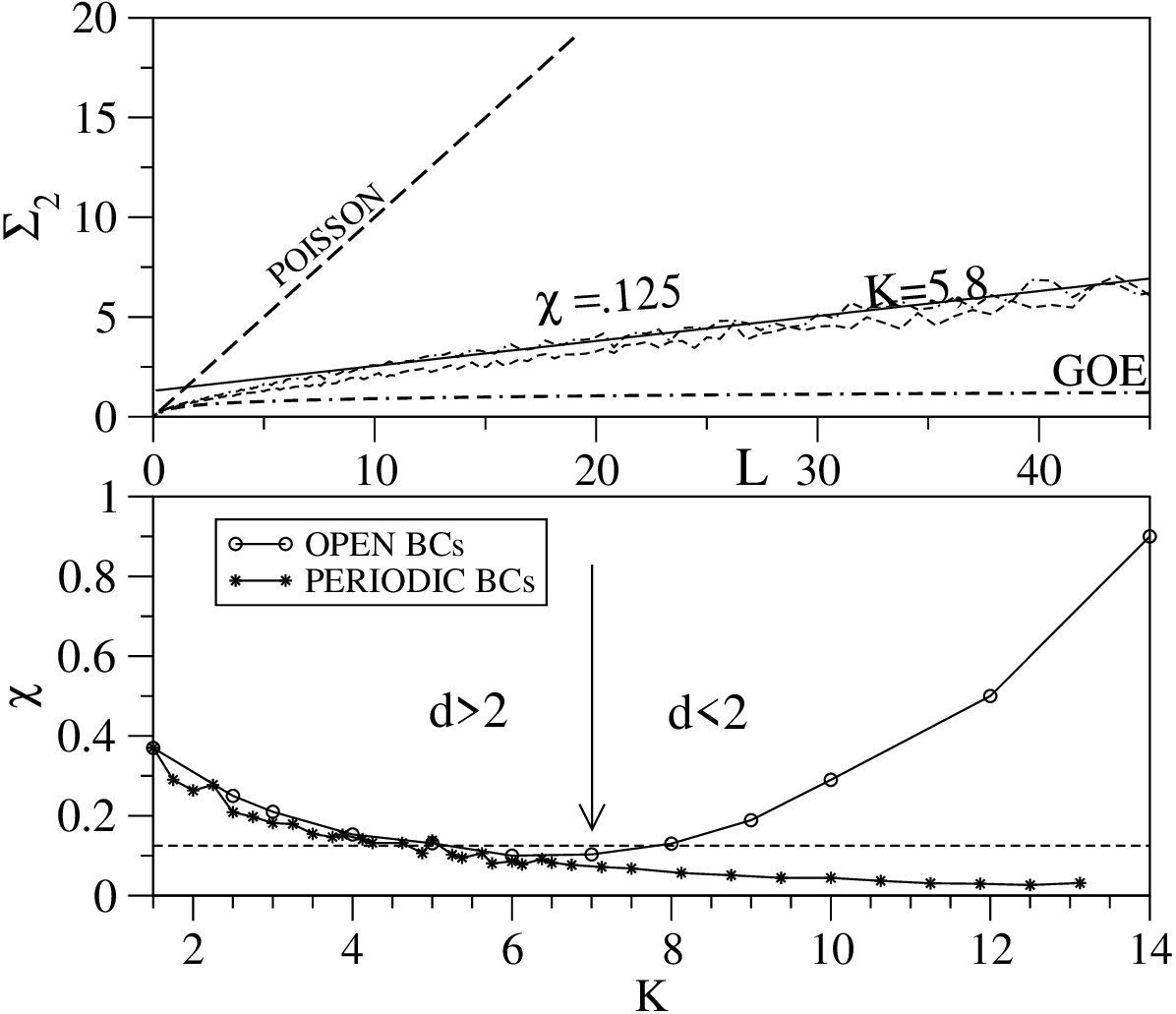}
\caption {Upper figure: shows that the variances 
($\Sigma_2$ statistics) for open BCs can pass close
to the so-called `critical' regime of near-linear variances, with a
slope close to $\chi \approx 0.125$.
Lower figure: compares values of $\chi$ for `open'
BCs (a unitary matrix of dimension $N_{tot}=10,000$ is diagonalized  
to obtain cells of dimension $N \approx 1250$) with `closed'
BCs (a finite matrix of dimension $N \approx 1250, \ M=51$ 
is diagonalized, with
periodic BCs to retain unitarity). The results are in
excellent agreement before delocalization $d>2$. After
delocalization $d<2$, the `closed' BCs tend to GOE, the
`open' BCs tend to Poissonian behaviour, providing a clear
signature of delocalization into multiple cells. }
\label{Fig13}
\end{figure}

We see here that $ \chi \approx 0.125= 1/2(1- \nu)$ would 
imply $\nu \approx 0.75$,
close to the exponent we obtained previously. This indicates
that here $\nu$, the dominant exponent near the golden-ratio
cantori, may play a role equivalent to the multi-fractal exponent $D_2$ in 
the Anderson MIT. This is a significant result as such behaviour
has not been seen and is not expected in a KAM system \cite{Verbaarschot}.

Unlike the billiard systems, here phase-space is not homogeneously filled
with cantori. However, underlying classical trajectories may spend 
considerable time 
trapped within the fractal trapping regions which border the cell. 
 The corresponding typical 
quantum states for $R > 1$ also sample the  cantori region,
but localize inhomogeneously in the fractal regions.
As $R$ increases, however, the distribution delocalizes 
into a more typically ergodic regime and the variances evolve towards
the GOE ($\Sigma_2(L) \sim Ln \ L$ ) limit.

\begin{figure}[h]
\includegraphics[width=3.in]{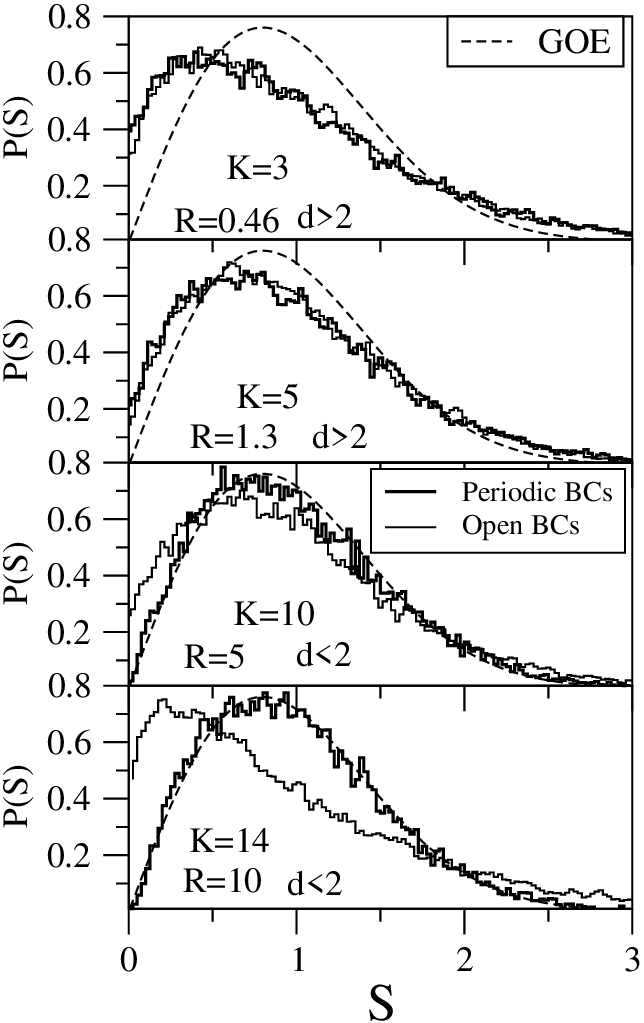}
\caption {Behaviour of NNS distributions equivalent
to the results in Figs.\ref{Fig12} and Fig.\ref{Fig13}. 
The results are insensitive
to BCs before the onset of delocalization onto multiple cells $d>2$. After
delocalization $d<2$, the `closed' BCs yield GOE, the
`open' BCs tend to more Poissonian behaviour. }
\label{Fig14}
\end{figure}

\section{OPEN BOUNDARY CONDITIONS: delocalization from single to multiple cells}

We diagonalize the matrix of $U^{\epsilon}$, with elements given
by Eq.\ref{eq32} in a basis of a given parity, dimension $N_{TOT}=10,000 \gg N$, for various
$K$,$\epsilon$ and $\hbar$ and obtain all eigenvalues and eigenvectors.
For our calculations the cell dimensions are in the range $N \simeq 600-2500$
and hence each diagonalization spans $N_{TOT}/N$ cells (fewer are
kept since states in the cells at the edges of the matrix are discarded).

For instance results presented below have $\epsilon=0.04$ and $\hbar=1/8$.
Hence, each cell contains $N=2\pi/(\hbar \epsilon) \simeq 1257$ states
so each diagonalization obtains up to 7 complete cells. 
Each full spectrum of length $N_{TOT}=10,000$ is now split into 
$8$ single cell sub-spectra. We assign the $i-th$ eigenvalue to the $m-$th cell if
\begin{equation}
(2m-1)\frac{\pi}{\epsilon}  \leq  \overline{p_i}  < 
\ (2m+1) \frac{\pi}{\epsilon} .
\label{eqbin}
\end{equation}
We calculate $P(S)$ and variances separately
for each of these sub-spectra of length $N = 1257$.
We then average the statistical distributions of 10-40 cells to obtain
smoother distributions.
We diagonalize
$U^{\epsilon}$ for basis states
between $l=10,000-20,000$ as easily as states between $70,000-80,000$,
for example, so as large a number of sets of 7 sub-spectra may be obtained as 
required. For the most delocalized spectra (with $K=14$)only a single 
central cell of about $1250$ states was sufficiently well converged
to be used for statistics.

Fig.\ref{Fig10} investigates how much of the momentum probability
for each eigenstate is contained in the cell it was assigned to: for $R \leq 1$
and $d >2$ the eigenstates are essentially fully contained within
the cells they are assigned to. However, at the onset of delocalization,
this procedure begins to fail; the expectation value $ \overline{p_i}$ of an 
eigenstate may assign it to the $m-$th cell; however, most of its
probability may in fact be trapped in neighbouring cells. This means that
increasingly, the eigenvalues of the sub-spectra become uncorrelated
and we can expect to see a return towards Poissonian statistics.

The apparent `failure' of this procedure, in fact provides a good
marker of delocalization from one-to-several cells. 
In contrast, the periodic boundary
conditions presented in the previous section, could
indicate only  the degree of `filling' of each single cell. 
Finally, we compare the spectral fluctuations for the two
different types of boundary conditions (periodic vs open) in
Figs.\ref{Fig12}-\ref{Fig14}. In Fig.\ref{Fig12} and Fig.\ref{Fig13} we show
that the spectral variances are insensitive to the boundary conditions
for $d>2$. However for $d<2$, the variances with periodic boundary
conditions gradually make a transition to GOE behaviour, while the
variances for the `open' system begin to return back to Poissonian statistics.
Fig.\ref{Fig14} shows that the NNS distributions closely follow the same
trends.

\section{CONCLUSIONS AND DISCUSSION}

In conclusion, we have investigated the quantum behaviour of atoms
exposed to pairs of $\delta$-kicks and shown that the cellular structure
arises from a novel oscillatory band structure of the corresponding
unitary matrix. One consequence is a new type of localization-delocalization
transition not seen in the QKR, where states delocalize from a single
cell to many, associated with a characteristic spectral signature (the return
to Poissonian statistics).

We have also found certain scalings $L \sim \hbar^{-0.75}$ and
$P(t) \sim t^{-0.75}$ which we argue result from the bands
of cantori present on the borders of each cell. We argue also that
the $0.75$ exponent may be identified with the dominant stability 
exponent obtained previously near golden-ratio cantori.
We show that the spectral fluctuations (both the NNS and spectral
variances) show important differences with the QKR in regimes 
where the delocalization of eigenstates is hindered by cantori.
The numerics provides evidence for behaviour somewhat
analogous to `critical' statistics of non-KAM billiards
 (in particular linear variances, with $\chi \approx 0.125$).
These results are novel  but important questions remain:
while the $L \sim \hbar^{-0.75}$ dependence might be explained
thus, it remains unclear why a similar exponent should also
be found in the  decay of the return probabilities $P(t) \sim t^{-0.75}$.
The question of whether the $\nu= 0.75$ exponent is somehow
equivalent to $D_2$ in the MIT and non-KAM billiards (in other
words whether $\nu$ related to any underlying multifractal character
of the wavefunctions) remains to be addressed in future.

By implication, the study shows that the behaviour
of cold atoms in double-pulsed standing waves of light is quite different
from the single-pulsed systems. Some aspects were already identified in
the experiments of \cite{Jones} and may have applications in atom optics
and atom chips, possibly as a mechanism for selecting atoms according 
to their momentum.

\begin{acknowledgments}
We are grateful to Phil Jones for the use of the
experimental data in Fig.\ref{Fig1}. CEC and TM acknowledge support
from the Engineering and Physical Sciences Research Council.
 We acknowledge the hospitality of the organisers of the workshop on
"Resonances and Periodic Orbits: Spectrum and Zeta Functions in Quantum
and Classical Chaos" at Institut Henri Poincare, Paris, in 2005. SF would
like to thank Richard Prange and Edward Ott for the hospitality at the
University of Maryland, where the work was completed.
\end{acknowledgments}

\appendix

\section{Derivation of the approximation (Eq.\ref{eq34}) 
for $U^{\epsilon}_{lm}$ }
 
In this Appendix Eq.\ref{eq34} will be derived for small
$\epsilon$. The crucial point in this derivation is that
contributions to the sum in Eq.\ref{eq32} of terms where
\begin{equation}\label{A1}
|k-l|\gg K/\hbar \qquad |k-m|\gg K/\hbar
\end{equation}
are negligible. This implies that for terms that contribute
appreciably, $k$ is close to $l$ and $m$. Therefore the
corresponding operator $\hat{k}=\hat{p}/\hbar$ can be written in
the form
\begin{equation}\label{A2}
\hat{k}=k_0+\hat{k}_1
\end{equation}
where $k_0$ is a number close to $l$ and $m$. Appreciable
contributions to the sum are found only for
\begin{equation}\label{A3}
|k_1| \leq K/\hbar.
\end{equation}
If $\hbar\epsilon(K/\hbar)^2 \ll 1$, as is the case for small
$\epsilon$, the condition $\hbar\epsilon k_1^2 \ll 1$ is
satisfied, justifying the approximation
\begin{equation}\label{A4}
\exp{-i\frac{\hat{p}^2\epsilon}{2\hbar}}=\exp{-i\frac{\hbar\epsilon\hat{k}^2}{2}}\approx
\exp{-i\frac{\hbar\epsilon k_0^2}{2}}\exp{-i\hbar\epsilon
k_0\hat{k}_1}.
\end{equation}
Substitution in Eq.3.1 with
$\hat{k}_1=-i\frac{\partial}{\partial x}$, and making use of the
fact that $\exp{-i \hbar\epsilon k_0 \frac{\partial}{\partial x}}$
is the shift operator, one finds
\begin{eqnarray}
U^{\epsilon}_{lm} \approx e^{-i\frac{\hbar}{2}[(T-\epsilon)l^2
+\epsilon k_0^2]}\\ \nonumber
 \langle l | e^{-i\frac{K}{\hbar}[\cos{x}
+\cos{(x-\hbar\epsilon k_0)]}}e^{-i\hbar\epsilon k_0\hat{k}_1}|m
\rangle,
\label{A5}
\end{eqnarray}
that reduces to
\begin{eqnarray}
U^{\epsilon}_{lm} \approx e^{-i\frac{\hbar}{2}[(T-\epsilon)l^2
+\epsilon k_0^2+2\epsilon k_0 m]} \\ \nonumber
\langle l |
e^{-i\frac{2K}{\hbar}[\cos{(x-\frac{1}{2}\hbar\epsilon
k_0)}\cos{(\frac{1}{2}\hbar\epsilon k_0)}]} |m \rangle.
\label{A6}
\end{eqnarray}
The matrix element is calculated with the help of the identity
\begin{equation}\label{A7}
\frac{1}{2\pi}\int_{-\pi}^{\pi}dx e^{-i\beta
\cos{(x-\alpha)}}e^{imx}=e^{im(\alpha-\pi/2)}J_m(\beta)
\end{equation}
resulting in
\begin{eqnarray}
U^{\epsilon}_{lm} \approx e^{-i\frac{\hbar}{2}[(T-\epsilon)l^2
+\epsilon k_0^2+\epsilon k_0 m+\epsilon k_0 l]}\\ \nonumber
e^{i\frac{\pi}{2}(l-m)}
J_{m-l}\left(\frac{2K}{\hbar}\cos{(\frac{1}{2}\hbar\epsilon
k_0)}\right).
\label{A8}
\end{eqnarray}
Indeed appreciable contributions are found only for $|m-l|\lesssim
2K/\hbar$ in agreement with (\ref{A1}). Since $k_0$ in close to
$l$ and $m$, within the approximation of this Appendix, it can be
replaced by one of these. The substitution $k_0=l$ results in
Eq.\ref{eq34}.

\end{document}